\documentclass{elsart}

\usepackage{graphics}
\usepackage{graphicx}
\usepackage{epsfig}
\usepackage{amssymb}

\begin{document}
\def\kket{\rangle \mskip -3mu \rangle}
\def\bbra{\langle \mskip -3mu \langle}

\def\ket{\rangle}
\def\bra{\langle}

\def\pard{\partial}

\def\sinh{{\rm sinh}}
\def\sgn{{\rm sgn}}

\def\alp{\alpha}
\def\del{\delta}
\def\Del{\Delta}
\def\eps{\epsilon}
\def\gam{\gamma}
\def\sig{\sigma}
\def\kap{\kappa}
\def\lam{\lambda}
\def\ome{\omega}
\def\Ome{\Omega}

\def\th{\theta}
\def\vphi{\varphi}

\def\Gam{\Gamma}
\def\Ome{\Omega}

\def\kav{{\bar k}}
\def\vb{{\bar v}}

\def\abf{{\bf a}}
\def\cbf{{\bf c}}
\def\dbf{{\bf d}}
\def\gbf{{\bf g}}
\def\kbf{{\bf k}}
\def\lbf{{\bf l}}
\def\nbf{{\bf n}}
\def\pbf{{\bf p}}
\def\qbf{{\bf q}}
\def\rbf{{\bf r}}
\def\ubf{{\bf u}}
\def\vbf{{\bf v}}
\def\xbf{{\bf x}}
\def\Cbf{{\bf C}}
\def\Dbf{{\bf D}}
\def\Kbf{{\bf K}}
\def\Pbf{{\bf P}}
\def\Qbf{{\bf Q}}

\def\omet{{\tilde \ome}}
\def\gammat{{\tilde \gamma}}
\def\Ft{{\tilde F}}
\def\ut{{\tilde u}}
\def\bt{{\tilde b}}
\def\vt{{\tilde v}}
\def\xt{{\tilde x}}

\def\ph{{\hat p}}

\def\vt{{\tilde v}}
\def\wt{{\tilde w}}
\def\phit{{\tilde \phi}}
\def\rhot{{\tilde \rho}}
\def\Ft{ {\tilde F}}
\def\mut{ {\tilde \mu}}

\def\Cb{{\bar C}}
\def\Nb{{\bar N}}
\def\Ab{{\bar A}}
\def\Db{{\bar D}}
\def\etab{{\bar \eta}}
\def\gb{{\bar g}}
\def\nb{{\bar n}}
\def\bb{{\bar b}}
\def\Pib{{\bar \Pi}}
\def\rhob{{\bar \rho}}
\def\phib{{\bar \phi}}
\def\psib{{\bar \psi}}
\def\omeb{{\bar \ome}}

\def\Sh{{\hat S}}
\def\Wh{{\hat W}}

\def\SS{I}
\def\psiw{{\xi}}
\def\tI{{g}}

\def\Ep#1{Eq.\ (\ref{#1})}
\def\Eqs#1{Eqs.\ (\ref{#1})}
\def\EQN#1{\label{#1}}

\newcommand{\beqa}{\begin{eqnarray}}
\newcommand{\eeqa}{\end{eqnarray}}
\newtheorem{mydef}{Definition}
\newtheorem{thm1}{Theorem}
\newtheorem{lem1}[thm]{Lemma}
\newenvironment{example}[1][Example]{\begin{trivlist}
\item[\hskip \labelsep {\bfseries #1}]}{\end{trivlist}}

\begin{frontmatter}



\title{Computability of entropy and information in classical Hamiltonian systems}


\author{Sungyun Kim}

\address{Asia Pacific Center for theoretical physics, Postech, San 31,
Hyoja-dong, Nam-gu, Pohang, Gyoungbuk, Korea, 790-784}
\ead{ksyun@apctp.org}
\begin{abstract}
We consider the computability of entropy and information in
classical Hamiltonian systems. We define the information part and
total information capacity part of entropy in classical Hamiltonian systems using
relative information under a computable discrete partition.
 Using a recursively enumerable nonrecursive set it is shown that even though the
initial probability distribution, entropy, Hamiltonian and its partial derivatives are
computable under a computable partition, the time evolution of its information capacity under the original
partition can grow faster than any recursive function. This implies that even though the probability
measure and information are conserved in classical Hamiltonian time evolution we might not
actually compute the information with respect to the original computable partition.
\end{abstract}

\end{frontmatter}

\section{Introduction}
\label{intro}
 In statistical mechanics, entropy is one of the most important
 concept.
  Entropy is the measure of uncertainty, and statistical
 mechanics can be viewed as the best theory we can get with the
 constraint of uncertainty or partial information \cite{Jaynes1, Jaynes2}. Even though
  the Hamiltonian time evolution conserves the probability and probability measure,
 the second law of thermodynamics states the entropy is nondecreasing in time.
 From the information theoretical perspective this may imply that
 information is lost during the computation of Hamiltonian time evolution.
 With rapid advances of numerical computation of physical systems,
  up to which extent we can actually compute the entropy and keep the information
  during its time evolution
 became an important and interesting problem.

 Computability means that there is an algorithm of calculating a
 quantity with a Turing machine up to arbitrary precision.
Originally this algorithm issue is related to the Hilbert's
plan to prove or disprove all statements derived from axiomatic systems
 in a systematic way. But
 in 1930s,  G\"odel showed that in
 an axiomatic system which is strong enough to express natural
 numbers there exist unprovable statements \cite{Goedel}.
Turing applied this to the programs and showed there exist
problems which are not solvable by algorithms \cite{Turing}. With the
advances of algorithmic information theory, G. Chaitin showed that
in a formal system with
 $n$ bits of axioms it is impossible to prove that a particular binary
 string is of Kolmogorov complexity greater than $n+c$ \cite{Chaitn}.

In analysis and differential equations what kind of quantities can be computed is also researched \cite{Aberth}.
In the wave equations and ordinary differential equations
Pour-El, I. Richards and Zong showed that
 a computable initial condition may evolve to a noncomputable solution at the later time
 \cite{PEl_p1,PEl_p2,PElZong}, and discussed how these phenomena would be related
 to the actual computation with Turing machine \cite{WeiZong}.
 The undecidability and computability in physical systems are also studied
 \cite{Kanter, Moore}.
C. Moore showed a Hamiltonian system can be mapped into a Turing
machine, and where the trajectories are passing can be mapped into
the Halting problem. Z. Xia \cite{Xia} showed that a gravitational
system may have non-collision singularity which makes the system not
computable. Noncomputability of topological entropy in various
systems are also researched \cite{Hurd,simonsen}. Recently D.
Gra\c{c}a et.al. \cite{Graca0, Graca} showed that an ordinary
differential equation can be mapped into a Turing machine.

  In this article we apply computability approach \cite{PourEl, Weihrauch} to
 the
 entropy and information of a probability distribution in classical Hamiltonian systems.
 We define the entropy of a continuous probability distribution through
 a discrete computable partition in the phase space.
 This entropy we define are divided as two parts, one representing
 information and the other representing information capacity.
 We show that even though the initial entropy and Hamiltonian is computable the time
 evolution of entropy may not be computable.

\section{Entropy, information and information capacity}

Let us first define entropy and information for the discrete probabilities
and probability distributions.
We follow Shannon's definition \cite{Shannon}.
Shanonn entropy for discrete probabilities $P_1,...,P_n,...$ is
 \beqa
 S = - \sum_i P_i \log P_i. \EQN{SD}
 \eeqa
 If we take
the base of the logarithm as $2$ then the unit of entropy is bits.
This entropy is a measure of uncertainty or the ability to store information. If
we have an unknown digit $X$ which can be either $0$ or $1$
  with probability $1/2$ each, we have 1 bit of entropy. The system has
  one bit of uncertainty or the capacity to store one bit of information.
 If the unknown digit $X$ is identified as $0$, the probability for $0$ is $1$ and probability
 for $1$ is $0$. From \Ep{SD} entropy becomes 0 ( $0\log 0$ is considered 0 as the limit value)
 and we say we
  gain 1 bit of information and entropy is reduced to 0 bit.

  Let us apply \Ep{SD} for continuous probability distribution
  inside a phase space $\Ome$ given by the momenta and coordinates
  $(\pbf, \qbf)$.
 Consider the probability distribution
  function $\rho (\pbf, \qbf,t)$ of a particle in the phase space, which satisfies Liouville's
  equation. Suppose that $\rho (\pbf, \qbf,t)$ has a finite support.
  To define probabilities and actually compute them continuous phase space
  and the probability distribution are discretized.
   Suppose that we discretize the phase space by countable number of cells with the
   same volume $\mu$.
   The probability distribution is also discretized by
   \beqa
 \rho_i  \equiv\frac{\int_i \rho d\Ome}{\mu}, \EQN{rhoi}
\eeqa
 where the integral at RHS of \Ep{rhoi} is over the $i$th cell.
 $\rho_i$ and $\mu$ satisfies the relation
\beqa
 \sum_i\rho_i \mu = 1. \EQN{Pnormal}
 \eeqa

 Using Shannon's definition, the entropy of the system $S (\rho, \mu)$ is
 \beqa
  & &S(\rho, \mu) = -\sum_i P_i \log P_i
  =- \sum_i \rho_i \mu \log \rho_i \mu \nonumber \\
  & &=-\sum_i \rhot_i \mut \log \rhot_i  -\sum_i \rhot_i \mut \log \mut
  = -\sum_i \rhot_i \mut \log \rhot_i  - \log \mut
  \EQN{Srm1}
 \eeqa
 where $\rhot_i = \rho \alp$ and $\mut = \mu/\alp$.
 The scaling constant $\alp$ is inserted to make the argument of $\log$ dimensionless.
  As the cell size $\mu$ becomes smaller and smaller, the first term in RHS (right hand side) of
  \Ep{Srm1}
  converges to the phase space integral $- \sum_i \rho_i \log (\rho_i \alp) \mu$
   $\rightarrow$ $-\int d\Ome \, \rho \log (\rho\alp)$ if the integral exists. The second term diverges
  as $\log (\alp / \mu)  $. These terms can be interpreted as follows.

If $\alp$ is chosen as the volume of probability distribution's support,
 the first part in  \Ep{Srm1} is always nonpositive.
 This term is called negative of so-called relative entropy or
 Kullback-Leibler divergence \cite{Kullback}. In general the Kullback-Leibler
 divergence is defined for two probability distributions $P(x)$ and $Q(x)$ as
 $D_{KL} (P||Q) = \int P(x) \log( P(x)/Q(x) ) dx$ and always nonnegative.
 For our choice of $\alp$ the first term in integral limit is always nonpositive and becomes zero only when the probability distribution is uniformly distributed on its support. Since the negative entropy decreases uncertainty it can be interpreted
 as the information we gain for the probability distribution $\rho$ with respect to the
 uniform distribution $1/\alp$. This term is also referred as 'physical entropy \cite{Latora}' or
 'coarse-grained entropy' over measure \cite{Matyas}. In the integral limit (often setting
 $\alp=1$) $\int d\Ome \rho \log (1/\rho)$ is called a 'differential entropy'.
The second term is the total information capacity, which is
positive when $\alp / \mu  >1 $. This term depends on how
fine the measures are.
   The information is minimum when $\rho$
 is uniform distribution $1/\alp$ on the support $\alp$. In descrete
 case the information is maximum $\log N$ when $\rho$ is discrete delta
  ($\rho = N$ for one cell and zero for all others), and diverges
  as the distribution goes to the Dirac delta distribution. This coincides with the fact that
  in general infinite digits are needed to specify a real number.
   The entropy and information we defined in \Ep{Srm1} are
subjective. First they depend
 on  the chosen partition, in this case discrete grid size
$\mu$ which determines the coarse grained precision. Second they depend on the scale $\alp$,
which determines how the we divide the information part and information capacity part.

 With this definition of entropy and information we consider the
 entropy of probability distribution with
Hamiltonian time evolution
with respect to the initial discrete partition.
 Since it is a classical Hamiltonian system,
 as time changes the probability density moves like an incompressible
 fluid in phase space, i.e. if one follows the time evolution of a point in
 phase space the density at the representation point remains
 constant and the measure is conserved. In actual computation of time evolution
 of initial probability distribution, one first discretize the probability density and
 evolve this discretized probability density with time (in most cases the probability
 distribution at later times can be only known numerically).

\begin{figure}[htb] 
\begin{center}
\includegraphics[height=1.8in, width=4.in]{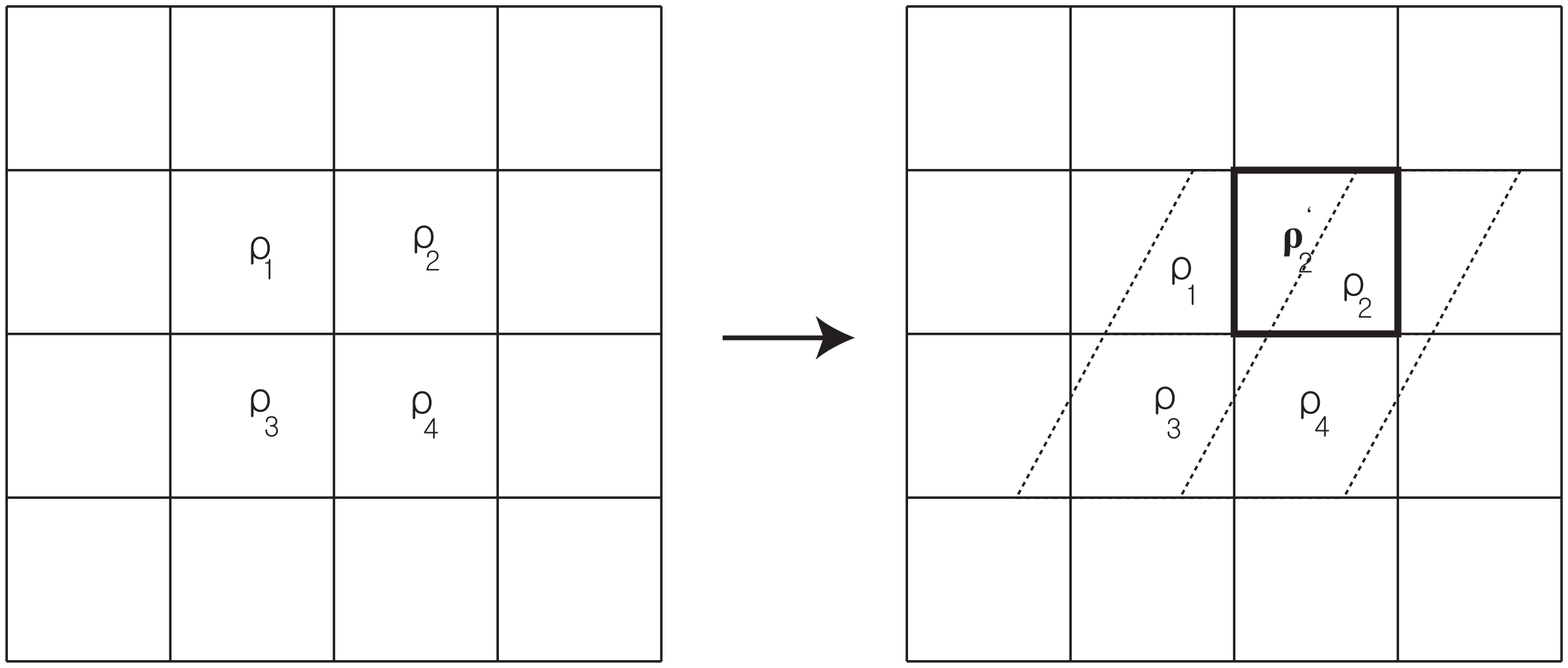}
 \caption{The new discretized probability distribution $\rho_i (t)$. In the left figure, each square shaped cells
 has discretized probability density $\rho_i (0)$s. ($i=1,..,4$) After one discrete time step the cells
 are deformed (shown as dashed parallelograms). The new discretized probability density
 $\rho_i (t)$ in $C_i (0)$ cell (the square with thick line in the right figure)
 is obtained by averaging the portions of probability densities moved into the $C_i (0)$ cell.} \label{phasecell}
\end{center}
\end{figure}

  Suppose that at time $t=0$ we make a discrete partition of phase
  space. This discrete partition divides the phase space with countable
  number of cells. Let us call
 the initial $i$th discretized probability density
 and $i$th partitioning cell as $\rho_i (0)$ and
 $C_i (0)$, like in \Ep{rhoi}. Then we compute the time evolved, discretized 
 probability density. In most cases the exact analytic form of $\rho (t)$ is not known,
 so the discretized probability in time $t$ is obtained by the time evolution
 of $\rho_i(t)$, which is the discretized probability density at $t=0$. 
 As time passes the original cell $C_i (0)$ deforms to $C_i (t)$, but the
 discretized probability distribution inside the deformed cell is still $\rho_i (0)$.
 The deformed cell $C_i (t) $ will be spread over the original discretized partition.
 If our computing precision is fixed, then the new discretized probability density
 is computed by the same discretized cells at $t=0$. Let us write the new
 discretized probability distribution within the initial $i$th cell
 as $\rho_i (t)$ (see Fig.~\ref{phasecell}).
  We have
  \begin{eqnarray}
  P_{i}(t)= \rho_i (t) \mu = \sum_j a_{ij} (t) \rho_j (0) \mu \EQN{Pi1}
 \eeqa
  where $a_{ij}(t)$ is given by
  \beqa
  a_{ij} (t)=\frac{ \mbox{volume of $C_i (0) \cap C_j (t)$} }{\mbox{volume of $C_i (0)$}}. \EQN{adef}
  \end{eqnarray}
  This probability distribution averaging (coarse graining) is the place where
  the information is lost due to the finite information capacity.

 From \Ep{adef} we have the relation
 \begin{equation}
 0 \le a_{ij} (t)  \le 1, \;\;\;\sum_i a_{ij}(t)=\sum_j a_{ij} (t)= 1. \EQN{aji}
 \end{equation}
  After one time step, the new
 entropy $S (t)$ computed with $P_i (t)$ using fixed precision $C_i (0)$ is
 \begin{eqnarray}
  & &S (t) = -\sum_i P_i(t) \log P_i(t)=  -\sum_i \sum_j a_{ij}(t)
  \rhot_j (0) \mut \log \bigg( \sum_{j'}
  a_{ij'}(t)\rhot_{j'} (0) \mut \bigg) \nonumber \\
  & &= -\sum_i \sum_{j} a_{ij}(t)
  \rhot_j (0) \mut \log \bigg( \sum_{j'}
  a_{ij'}(t)\rhot_{j'} {(0)} \bigg) -\log \mut
 \EQN{H1}
  \end{eqnarray}
  where \Ep{Pnormal} and \Ep{aji} is used.
 Since the function $f(x) =x \log (\lam x)$ with $\lam>0$ is a convex
 function and the convex function satisfies Jensen's inequality
  \begin{equation}
   f( \sum_i a_i x_i) \le \sum_i a_i f(x_i)\;\;\; \mbox{for all $a_i \ge 0$},\EQN{Jensen}
   \end{equation}
we have
 \beqa
& &-\sum_i \sum_j  a_{ij} (t) \rhot_j (0) \mut \log \bigg( \sum_{j'} a_{ij'} (t) \rhot_{j'} (0) \bigg)  \nonumber \\
& &\ge -\sum_i \sum_j a_{ij} (t)  \rhot_j (0)  \mut
\log \rhot_j (0) = - \sum_i  \rhot_i (0)  \mut \log \rhot_i (0)
\EQN{Sineq}
 \eeqa
 We see that the information capacity part of $S (t)$ in \Ep{H1} is the same, but
  the negative entropy (information) part of $S (t)$ is
 always  greater or equal than the information part of
 $S (t)$. This means that the information is always same or lost due to the
 coarse graining. One way to avoid
 the information loss is using finer partition. If the entropy is
 calculated for each fractional probabilities of $P_{ij}(t)=a_{ij} (t) \rhot_j (0) \mut$,
 we have
 \beqa
& &S_{finer} (t)= -\sum_{j} \sum_{i} P_{ij}(t) \log P_{ij}(t)=    -\sum_{j} \sum_{i} a_{ij}(t) \rhot_j (0) \mut \log
(a_{ij}(t) \rhot_j (0) \mut) \nonumber \\
& &= -\sum_{j} \sum_i a_{ij} (t) \rhot_j (0) \mut \log (a_{ij} (t) \rhot_j (0) \mut) \nonumber \\
& & = -\sum_{i} \rhot_i (0) \mut \log \rhot_i (0) -\log \mut +
\sum_{i} \rhot_i (0)  \mut \bigg( - \sum_{j}  a_{ij} (t) \log a_{ij} (t) \bigg)
\EQN{S1f}
 \eeqa
 The first term in RHS of \Ep{S1f} is information, which is the same
 as before. But the total information capacity is increased by
 \beqa
\sum_{i}  \rhot_i (0)  \mut \bigg( - \sum_{j}  a_{ij} (t) \log a_{ij} (t) \bigg), \eeqa
 where $- \sum_{j}  a_{ij} (t) \log a_{ij} (t)$, which is always nonnegative since $a_{ij} (t)\ge 0$,
  represents the entropy increase (information capacity increase)
 due to the finer partition or resolution. For example in baker transformation this
 term is $\log 2=1$ bit for each discrete time step.

 Now we ask the question of computation of entropy and its time evolution. As stated before, the total
 entropy is a subjective quantity which depends on the partition we choose. Given the computable partition and computable initial conditions,
  can we compute the information and the total information capacity needed to keep the information during time evolution?  To answer this question, we first need definitions
 about computability.

\section{Computability preliminaries}
\label{compre}

 The following definitions, theorems and examples are from the book of
 M. B. Pour-El and J. I. Richards \cite{PourEl}.
Here the term {\it recursive function} means it can be implemented
and calculated by a Turing machine. $\mathbb{N}$ denotes the set of
non-negative integers.

\begin{mydef}
A sequence $\{r_k \}$ of rational numbers is {\bf computable} if
there exists three recursive functions $a, b, s$ from $\mathbb{N}$
such that
$$r_k = (-1)^{s(k)} \frac{a(k)}{b(k)+1}\;\;\;\mbox{for all $k$} $$
\end{mydef}

\begin{mydef}
A sequence $\{r_k \}$ of rational numbers {\bf converges
effectively} to a real number  $x$ if there exists a recursive
function $e:\mathbb{N}\rightarrow \mathbb{N}$ such that for all $N$:
$$ k \ge e(N)\;\; \mbox{implies}\;\;\; |r_k-x|\le 2^{-N}$$.
\end{mydef}

\begin{mydef}
A real number $x$ is {\bf computable} if there exists a computable
sequence $\{r_k \}$ of rationals which converges effectively to $x$.
\end{mydef}

We now define computable functions. For simplicity we first consider
the case where the function $f$ is defined on a closed bounded
rectangle $I^q$ in $R^q$. Specifically $I^q = \{ a_i \le x \le b_i ,
1\le i \le q\}$ is called computable rectangle if $a_i$ and $b_i$
are computable reals.
\begin{mydef}
Let $I^q \subseteq \mathbb{R}^q$ be a computable rectangle. A
function $f: I^q \rightarrow \mathbb{R}$ is computable if:

(i) $f$ is {\bf sequentially computable}, i.e. $f$ maps every
computable sequence of points $x_k \in I^q$ into a computable
sequence $\{ f(x_k) \}$ of real numbers;

(ii) $f$ is {\bf effectively uniformly continuous }, i.e. there is a
recursive function $d:\mathbb{N} \rightarrow \mathbb{N}$ such that
for all $x,y \in I^q$ and all $N$:

  $|x-y|\le 1/d(N)$ implies
$|f(x)-f(y)| \le 2^{-N}$. \label{fcom1}

\end{mydef}

 Now the function in $\mathbb{R}^q$ is considered.

\begin{mydef}
A sequence of functions $f_n: \mathbb{R}^q \rightarrow \mathbb{R}$
is {\bf computable} if:

 (i) for any computable sequence of points $x_k \in \mathbb{R}^q$,
 the double sequence of reals $\{ f_n (x_k) \}$ is computable;

 (ii) there exists a recursive function
 $d:\mathbb{N}\times\mathbb{N} \times \mathbb{N} \rightarrow
 \mathbb{N}$ such that for all $M,n,N$:

 $|x-y| \le 1/d(M,n,N)$ implies $|f_n(x)-f_n(y)|\le 2^{-N}$ for all
 $x,y\in I^q_M$,

 where $I^q_M = \{ -M \le x_i \le M, 1\le i\le q \}$. \label{funcomp}
\end{mydef}

Theorems about computability of integrals of functions.

\begin{thm1}
Let $I^q$ be a computable rectangle in $\mathbb{R}^q$, and let $f_n: I^q \rightarrow
\mathbb{R}$ be a computable sequence of functions. Then the definite integrals
$$ v_n = \int ...\int_{I^q} f_n (x_1,...,x_q) dx_1...dx_q $$
form a computable sequence of real numbers. \label{int1}
\end{thm1}

\begin{thm1}
Let $f$ be a computable function on a computable interval $[a,b]$. Then the indefinite
integral
$$\int_a^x f(u) du$$
is computable on $[a,b]$. \label{int2}
\end{thm1}

Now we define the recursively enumerable nonrecursive set.

 A set $A \subseteq \mathbb{N}$ is called {\it recursively enumerable}
 if $A=\varnothing$ or $A$ is the range of a recursive function $a$.
 In other words, we can compute $a(0), a(1), a(2)...$ step by step
 using a Turing machine.

  A set $A \subseteq \mathbb{N}$ is called {\it recursive } if both
  $A$ and its complement $\mathbb{N}-A$ are recursively enumerable.

A fundamental and important theorem of logic is that
\begin{thm1}
There exists a set $A \subseteq \mathbb{N}$ which is recursively
enumerable but not recursive. \label{reNre}
\end{thm1}
If a set $A$ is recursively enumerable but nonrecursive, then we
have a recursive (or computable) procedure to get the elements
$a(0), a(1), a(2),...$ sequentially, but we have no recursive (or
computable) procedure to tell an arbitrary number $\alp \in
\mathbb{N}$ belongs to $A$ or not. We do not know how long we should
compute the sequence to see $\alp$ appears. This is expressed in the
following lemma.
\begin{lem1}
(Waiting lemma). Let $a:\mathbb{N}\rightarrow\mathbb{N}$ be a one to
one recursive function generating a recursively enumerable
nonrecursive set $A$. Let $w(n)$ denote the "waiting time"
$$w(n)=max\{m: a(m)\le n\}.$$
Then there is no recursive function $c$ such that $w(n)\le c(n)$ for
all n. \label{waiting}
\end{lem1}
 One example of recursively enumerable nonrecursive set is the set
of Halting programs.

Next theorem is about the convergence of sequence of functions. The
proof is in \cite{PourEl}.
\begin{thm1} (Closure under effective uniform convergence)
Let $f_{nk}:I^q \rightarrow \mathbb{R}$ be a computable double
sequence of functions such that $f_{nk}\rightarrow f_n$ as
$k\rightarrow \infty$, uniformly in $x$, effectively in $k$ and $n$.
Then $\{ f_n \}$ is a computable sequence of functions.
\label{Ceffec}
\end{thm1}

Now we show an example of a function which is not bounded by any recursive function,
which will be used later in section 4.
\begin{example}
 Let $a:\mathbb{N}\rightarrow\mathbb{N}$ be a one to
one recursive function generating a recursively enumerable
nonrecursive set $A$. We assume $0 \notin A$. Then
the function $f(z) = \sum_{m=0}^{\infty} z^m /a(m)^m$ is an entire function but
not bounded by any recursive function.
\label{exf}
\end{example}
 In this example, the sequence of Taylor coefficients $\{ 1/a(m)^m \}$ is
 computable. And this series is uniformly convergent on any
 compact disk $\{ |z| \le M\}$ where $M$ is a positive integer. To see this
 we note that there are only finitely many values of $a(m)$ with $a(m) \le M$.
 For all other $a(m) \ge M+1$, and inside the disk of $\{ |z| \le M\}$
 sum of the other terms containing only $a(m) \ge M+1$ are bounded by
 $\sum M^m/(M+1)^m$. So $f$ is uniformly convergent on any
 compact disk $\{ |z| \le M\}$.

  But the sequence of values $f(0),f(1),f(2),...$ are not bounded by any recursive function.
  For positive real argument $f(x)$ is larger than any single term in its Taylor series.
  For one term $m= w(n)$ where $w(n)$ is the waiting function of the sequence $a(n)$,
  We have
  \beqa
  f(2n) > \bigg( \frac{2n}{a(m)} \bigg)^m \ge \bigg( \frac{2n}{n} \bigg)^m = 2^m
  = 2^{w(n)} > w(n).
  \eeqa
  Hence $f(2n) > w(n)$ and $w(n)$ is not bounded by any recursive function, so $f(z)$ is not bounded
  by any recursive function.

With these preliminaries, next section we construct an example
in which the time evolution of entropy is not computable.

\section{Computability of time evolution of entropy}
 In this section we construct a Hamiltonian system, in which
 the Hamiltonian and its partial derivatives, initial probability distribution and information are computable
under a computable partition but the time evolution of entropy under the original partition
grows faster than any recursive function.

\begin{figure}[ht]
 \begin{minipage}[b]{.5\linewidth}
 \centering
 \includegraphics[scale=.5]{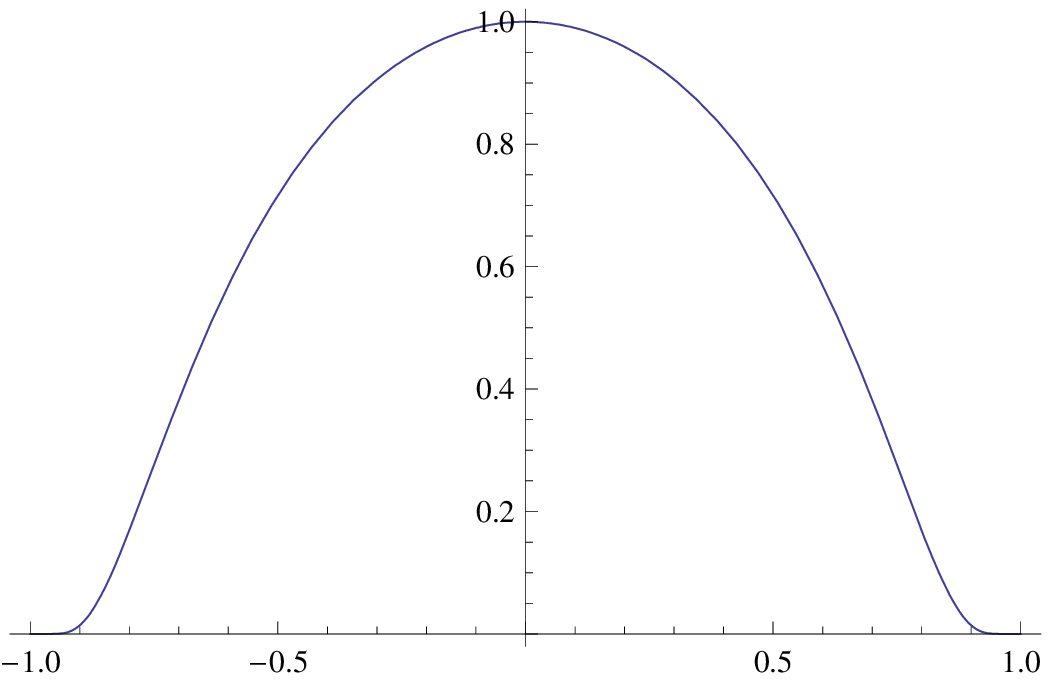}
 \caption{Plot of $\phi(x)$. $\phi(x)$ is a $C^{\infty}$ function with support $[-1,1]$.}
 \label{phiplot1}
 \end{minipage}
 \hspace{.5cm}
 \begin{minipage}[b]{.5\linewidth}
 \centering
 \includegraphics[scale=.5]{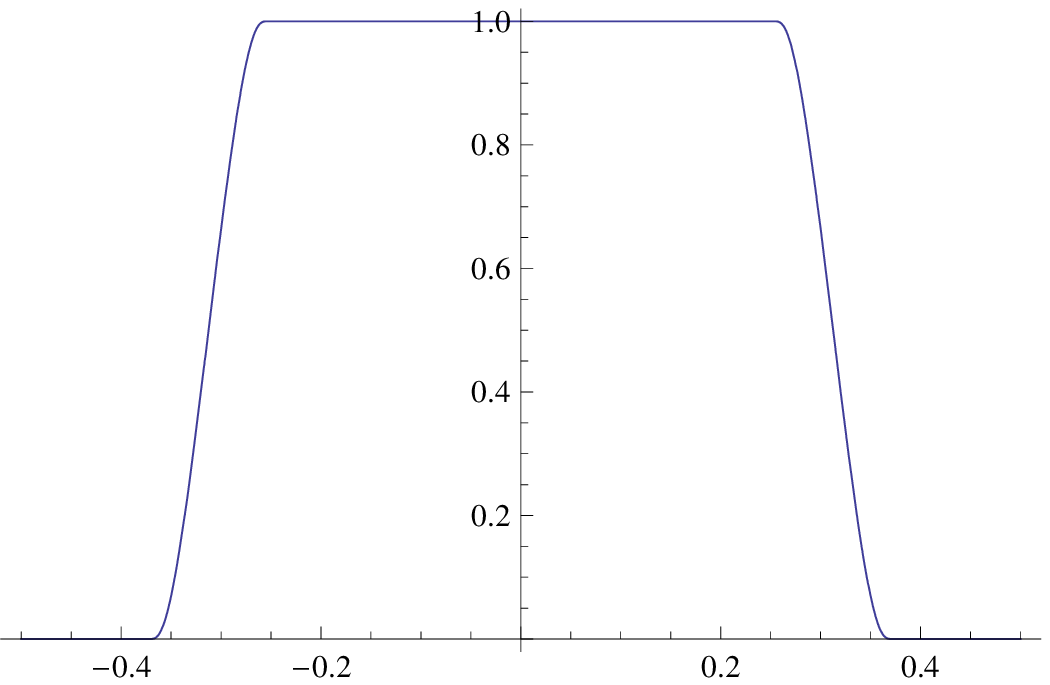}
 \caption{Plot of $\psi(x)$. $\psi(x)$ is a $C^{\infty}$ function with support $[-3/8,3/8]$.
 $\psi (x)$ has constant value 1 between $-1/4 < x <1/4$.}
 \label{psiplot1}
 \end{minipage}
 \end{figure}

 To construct our Hamiltonian and probability distribution
we first define a pulse function
 \beqa
 \phi(x) = \left\{ \begin{array}{ll} e^{-x^2/(1-x^2)}\;& \mbox{for $-1 < x <
 1$,} \\
 0. \;& \mbox{otherwise} \end{array} \right.
 \eeqa
  This function is in $C^{\infty}$ and has the support $[-1,1]$ (Fig.~\ref{phiplot1}).
  We define the normalization constant of $\phi(x)$ as
  $N_{\phi}=1/ \int_{-1}^{1} \phi(x) dx =0.828569... \;$. $N_{\phi}$
  is computable, since it is an integral of computable function under a bounded interval by theorem~\ref{int1}.

  We define another pulse function
 \beqa
 \psi(x) = \left\{ \begin{array}{ll} \phi_{int} (2^{4}(x+5/16))\;& \mbox{for $x< 0$,} \\
  \phi_{int} (-2^{4}(x-5/16)) \;& \mbox{for $0 \le x$}
  \end{array} \right.
  \eeqa
  where $\phi_{int}(x)$ is given by
  \beqa
  \phi_{int}(x) = \left\{ \begin{array}{ll} N_{\phi} \int_{0}^x \phi(x) dx \; & \mbox{for $x \ge -1$,}\\
  0 \;& \mbox{for $x<-1$.} \end{array} \right.
 \eeqa
  $\phi_{int}(x)$ is $0$ for $x<-1$, increases smoothly ($C^{\infty}$ way) from
  $0$ to $1$ for $-1\le x\le 1$ and $1$ for $1\le x$. This is a computable function by theorem~\ref{int2}.
  The shape of $\psi(x)$ is shown in Fig.~\ref{psiplot1}. It is a $C^{\infty}$ function, which
  increases from 0 to 1 in $C^{\infty}$ way on the interval $(-3/8,-1/4)$,
  constant value 1 between
$-1/4$ and $1/4$, and decreases from 1 to 0 in $C^{\infty}$ way on the interval
$(1/4,3/8)$. Otherwise it is 0.
  Both $\phi(x)$ and $\psi(x)$ are computable functions.

 Using above pulse functions $\phi(x)$ and $\psi(x)$ we construct
 the Hamiltonian and probability distribution function.
  We consider the 6 dimensional phase space, $(\pbf, \qbf) = (p_1,p_2,p_3, q_1,q_2,q_3)$.

   The Hamiltonian is constructed by
   \beqa
   H = \sum_{m=0}^{\infty} H_m (\pbf, \qbf) \EQN{Htotal}
   \eeqa
where $H_m$ is defined as
 \beqa
  H_m =m(e^{p_2} p_1 q_1 - q_2 ) \psi(q_3 -m).
  \eeqa
  Since
  each support of $H_m$ are in $m-3/8 \le q_3 \le m+3/8$ (none of them overlap),
  for any computable point $(\pbf,\qbf)$ we can make a ball centered at that point with radius $2^{-N}$
  and for sufficiently large $N$ this ball contains at most one $H_m$, which is a computable
  function.
  With this fact and definition~\ref{funcomp} we see that $H$ is a computable function in $\mathbb{R}^6$.
  Also all $\frac{\partial H_m}{\partial p_i}$ and  $\frac{\partial H_m}{\partial q_i}$ are computable
  and we can choose a neighborhood of computable point which contains at most one nonzero
  derivatives of $H_m$. By the same logic
  all $\frac{\partial H}{\partial p_i}$ and  $\frac{\partial H}{\partial q_i}$ are computable.

\begin{figure}[htb] 
\begin{center}
\includegraphics[height=3in, width=4.in]{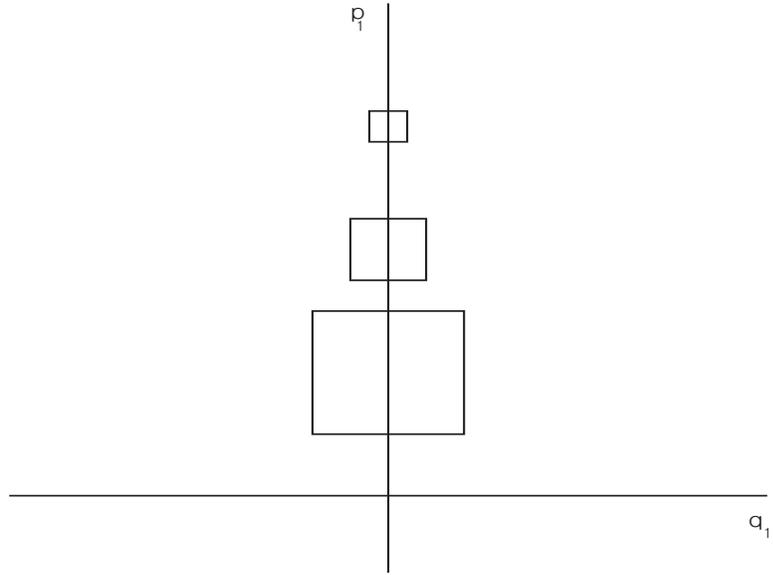}
 \caption{The supports of $\rho_{mn}$ for fixed $n$ with $m=1,2,3...$ in $p_1q_1$ space.} \label{rhomn}
\end{center}
\end{figure}

  The initial probability distribution is chosen as
  \beqa
  \rho (\pbf, \qbf) = N_{\rho} \sum_{n=0}^{\infty} \sum_{m=1}^{\infty} \rho_{mn} (\pbf, \qbf)
  \eeqa
  where
  \beqa
   & &\rho_{mn} =\frac{N_{\phi}^6}{ 2^{6n+m} a(n)^n} \phi (2^{n+m} (p_1 - m ))   \phi (2^{n+1} (p_2- 1/2   ))
   \phi (2^n p_3) \nonumber \\
   & &\times\prod_{j=1}^2 \phi (2^{n} q_j)  \phi (2^{n+2} (q_3-n)) \nonumber \\
   \EQN{rhomn2}
  \eeqa
   and $\{a(0),a(1),a(2)...\}$ is a recursively enumerable nonrecursive set from $\mathbb{N}$
   to $\mathbb{N}-\{0\}$.
  In \Ep{rhomn2}, the support of each $\rho_{mn} (\pbf,\qbf)$ is a 6 dimensional hypercube
  with three $2^{-n+1}$ length ($p_3,q_1,q_2$) sides, one $2^{-n}$ length $p_2$ side, one $2^{-n-1}$ length $q_3$ side and one $2^{-n-m+1}$ length $p_1$ side, centered
  at $(p_1,p_2,p_3,q_1,q_2,q_3) = (m,1/2,0,0,0,n)$.   (See Fig.~\ref{rhomn} )
  The probability inside each $\rho_{mn}$ is $\int d\pbf d\qbf \rho_{mn} = 2^{-12n-2m-3}/a(n)^n$.

 Again, each $\rho_{mn}$ is computable and  none of the supports of $\rho_{mn}$ overlap.
 For any computable point if we choose radius $2^{-N}$ sized ball centered at that point with large enough  $N$ it overlaps at most one nonzero $\rho_{mn}$ which is a computable function.
From definition~\ref{funcomp}  $ \rho (\pbf, \qbf)$ is computable.
 $N_{\rho}$ is the normalization constant, which makes $\int \rho ( \pbf, \qbf) d\pbf d\qbf=1$.
 $N_{\rho}$ is also a computable number since the sum $\sum_{m,n} \int d\pbf d\qbf \rho_{mn}
 = \sum_{m,n} 2^{-12n-2m-3}/a(n)^n $ converges faster than a geometric series.

 Next we compute the information and entropy of initial probability distribution.
 In \Ep{Srm1}, the entropy is defined through the probabilities
 with a computable partition. Here computable partition means that boundaries of the partition are
 made with computable functions and any computable finite area can be covered by increasing the number of partitions in a recursive way.
 Let us choose the original partition as
   $p_i = k 2^{-n_p}$ and $q_i= k 2^{-n_p}$ surfaces ($i=1,2,3$. $n_p$ is a (possibly large) natural number.
   $k=0,1,2,...$).
  Then the smallest cell in this partition is 6 dimensional hypercube with
  each side $2^{-n_p}$ and volume $\mu=2^{-6 n_p}$. For a given scale $\alp$ which
  defines the unit volume,
  $-\log (\mu/\alp)$ is the precision one can get for the volume and $-\log (2^{-n_p}/\alp)$
  is the precision for each coordinate. For now let us choose the scale $\alp$ as $1$.

  Under this partition, for any natural number $n_p$, the information part is
 $-\sum_i \rhot_i \mut_i \log \rhot_i < (6n+m)2^{-6n-m}$  for each $\rho_{mn}$ pulse.
 The total $\rho$ the information part is dominated by $\sum_{m,n} (6n+m)2^{-6n-m}$ which is
 effectively convergent, so the initial information under this partition is computable.
 The information capacity part is $ -\log \mut$
 and also computable.

    Now we consider the time evolution of probability distribution and its
    entropy and information under the original partition. Since the Hamiltonian time evolution is measure preserving,
    the term $-\sum_i \rhot_i \mut_i \log \rhot_i  $ is conserved during the time evolution
    if we make finer partition. But the information capacity term, which depends on
    the finer partition and ability to resolve the probability distribution with original
    partition
    , increases with time as
    the second term in RHS of \Ep{S1f} shows.

    The Hamiltonian equation is given by
    \beqa
    \dot{p_i} = -\frac{\partial H}{\partial q_i},\;\; \dot{q_i} = \frac{\partial H}{\partial p_i},
    \eeqa
    and the solution of the Hamiltonian in  \Ep{Htotal} is (Note that all $\rho_{mn}$ are between
    $n-1/4 \le q_3 \le n+1/4$ for any $n$.)
    \beqa
    & &\mbox{for $n-1/4 \le q_3 \le n+1/4$,} \nonumber \\
    & &p_1(t) = p_1(0) \exp( -e^{nt + p_2(0)} +e^{p_2(0)}),\;\;\; q_1(t) = q_1(0)\exp( e^{nt + p_2(0)} -e^{p_2(0)}), \nonumber \\
   & &p_2(t) = n t + p_2(0),\;\;\; q_2(t) = e^{p_2(0)}p_1(0)q_1(0) (e^{n t} -1)+ q_2(0), \nonumber \\ & &p_3(0)=p_3(0),\;\;\;q_3(t)=q_3(0).
    \eeqa
 This solution shows exponential of exponential squeezing and stretching in $p_1$ and $q_1$ directions.
 For example a rectangle in $p_1 q_1$ space with  side lengths $\delta p_1$ and $\delta q_1$
 at $t=0$ is stretched to a rectangle with side lengths $\delta p_1 \exp( -e^{n t+ p_2(0)}+e^{p_2(0)} )$
 and $\delta q_1 \exp(e^{n t +p_2(0)} -e^{p_2(0)} )$.

 If the partition in $p_1 q_1$ space is
 made with $\delta p_1 \delta q_1$ cells, then the time evolution of one cell
 is stretched and overlaps at least $N_s=[ \exp( e^{n t +p_2(0)} -e^{p_2(0)})]$ number of other cells
 in $q_1$ direction.
 ($[x]$ means the largest integer not larger than $x$.)  In $p_1$ direction $N_s$ number of partial cells
 are squeezed into one cell and at least $\log N_s$ bits of resolution is needed to distinguish
 the thin strips of cells in the original one cell.

 In view of the information capacity term
$\sum_{i} \rhot_i (0)  \mut \bigg( - \sum_{j}  a_{ji} (t) \log a_{ji} (t) \bigg)$,
for the $\rhot_i (0)$ which is in $n-1/4 \le q_3 \le n+1/4$
each $a_{ji} (t)$ is around $1/N_s $ and it is summed over $N_s$ terms. So
 \beqa
 - \sum_{j=1}^{N_s}  a_{ji} (t) \log a_{ji} (t) \approx \log N_s.
 \eeqa

  From \Ep{rhomn}, $0 \le p_2(0) \le 1$ for all $\rho_{mn}$ and
  \beqa
    \exp(e^{nt}) <  \exp( e^{n t +p_2(0)} -e^{p_2(0)}) < \exp (e^{2nt})
  \eeqa
  for $t \ge 2$.   If we consider the whole 6 domensional space the number of overlapping cells are larger
    due to the exponential  stretching in $q_2$ direction.
 So for each $\rho_i \mu$ segment which is in $n-1/4 \le q_3 \le n +1/4$
  the term $-\sum a_{ji}(t) \log a_{ji}(t)$ is bounded by
 \beqa
  e^{n t} \log e < -\sum_{j} a_{ij}(t) \log a_{ij}(t) < e^{2 nt } \log e
  \eeqa
 for $t \ge 2$.
  Considering all $\rho_{mn}$ pulses for fixed $n$, the information capacity increase in  $n-1/4 \le q_3 \le n +1/4$ is bounded by
 \beqa
 & &\sum_{m=1}^{\infty} 2^{-12n -2m-3 }  \frac{ e^{n t} \log e}{a(n)^n} =  \frac{\log e}{24} \bigg(
 \frac{2^{-12} e^t }{a(n)} \bigg)^n
     <  \nonumber \\
     & &\sum_{n-1/4 \le q_3 \le n +1/4 } \rho_i \bigg(-\sum_j a_{ij} (t) \log a_{ij} (t) \bigg)
 <   \frac{\log e}{24} \bigg(
 \frac{2^{-12} e^{2t} }{a(n)} \bigg)^n
 \eeqa
By summing over $n$ for the total probability distribution, we get
\beqa
 \frac{\log e}{24} \sum_{n=1}^{\infty}  \bigg(
 \frac{2^{-12} e^t}{a(n)} \bigg)^n
 <   \sum_{i} \rho_i \mu \bigg(-\sum_j a_{ij} \log a_{ij} \bigg)
 <\frac{\log e}{24}\sum_{n=1}^{\infty}   \bigg(
 \frac{2^{-12} e^{2t} }{a(n)} \bigg)^n.     \nonumber \\
 \eeqa
Like the example at the end of section~\ref{compre}, we see that the information capacity increase in time $t=0,1,2,3,...$ are finite but not bounded by any
 recursive function, for any $\mu$.
  The information capacity is related to the
 ability
 to describe how far a cell is stretched or how many other cells are squeezed into an original cell
 for probability pulses in $\mu$ accuracy.
But this grows faster than any recursive function and we cannot find a recursive way to compute
 information within the original computable partition.

\section{Summary}
 In summary, we defined information and information capacity in classical Hamiltonian
 system and showed an example in which the initial probability
 distribution and its information are computable,
 and the Hamiltonian and its derivatives are computable, but the
 information capacity increase is not bounded by any recursive function.
 Its total entropy, which is defined through a computable
 discrete partition, is originally computable but its time evolution
 grows faster than any recursive function. This total entropy
 is related to the precision required to compute the information,
 so the time evolution of information is not computable within the
 original computable discrete partition.
  Even though the information is a conserved quantity in the Hamiltonian
  time evolution, the result shows that we might not actually compute it. 

\ack
 The author would like to thank Moo Young Choi, Seunghwan Kim and
 Gonzalo Ordonez for helpful comments.



\pagebreak

\pagebreak

\pagebreak

\pagebreak

\end{document}